\documentclass[preprint,aps,prd,groupedaddress,showpacs,floatfix,%
nofootinbib]{revtex4}
\usepackage{dcolumn}
\usepackage{bm}
\usepackage{epsfig}
\usepackage{hyperref}
\usepackage{graphicx}
\usepackage{dcolumn}
\usepackage{bm}
\usepackage{longtable}
\begin{document}

\title{Semileptonic $B_c$ decays and Charmonium distribution amplitude}
\author{Tao Huang} \thanks{Email: huangtao@mail.ihep.ac.cn.}
\author{Fen Zuo} \thanks{Email: zuof@mail.ihep.ac.cn.}.
\affiliation{Institute of High Energy Physics, Chinese Academy of
Sciences, Beijing 100049, China}
\begin{abstract}
In this paper we study the semileptonic decays of the $B_c$ meson
in the Light-Cone Sum Rule (LCSR) approach. The result for each
channel depends on the corresponding distribution amplitude of the
final meson. For the case of $B_c$ decaying into a pseudoscalar
meson, to twist-3 accuracy only the leading twist distribution
amplitude (DA) is involved if we start from a chiral current. If
we choose a suitable chiral current in the vector meson case, the
main twist-3 contributions are also eliminated and we can consider
the leading twist contribution only. The leading twist
distribution amplitudes of the charmonium and other heavy mesons
are given by a model approach in the reasonable way. Employing
this charmonium distribution amplitude we find the cross section
$\sigma(e^+e^-\to J/\psi+\eta_c)\simeq22.8~\mbox{fb}$ which is
consistent with Belle and BaBar's data. Based on this model, we
calculate the form factors for various $B_c$ decay modes in the
corresponding regions. Extrapolating the form factors to the whole
kinetic regions, we get the decay widths and branching ratios for
various $B_c$ decay modes including their $\tau$ modes when they
are kinematically accessible.
\end{abstract}



\maketitle

\section{Introduction}
The $B_c$ meson has been observed by the CDF and D0 groups in
different channels \cite{CDF1,CDF2,D0}. The semileptonic decays of
$B_c$ was studied in Ref.~\cite{QM} using the BSW(Bauer, Stech,
Wirbel) model \cite{BSW} and the IGSW(Isgur, Grinstein, Scora,
Wise) model \cite{ISGW}, and in the frame work of the
Bethe-Salpeter equation in Ref.~\cite{BSE} and in the relativistic
constituent quark model in Ref.~\cite{RQM}. Alongside the small
differences in the partial decay widths in these models, the first
estimates made on the basis of the three points (3P) QCD sum rules
(SR) \cite{3PSR1} are significant smaller. The reason was supposed
to be the valuable role of Coulomb corrections, which implied the
summation of $\alpha_s/v$ corrections significant in the $B_c$
\cite{3PSR2}. It is suggested that the discrepancy observed
between the QCD sum rules and the quark models can be eliminated
by including these higher QCD corrections.

However, the 3PSR inherits some problems when describing
heavy-to-light transitions, the main one being that some of the
form factors have a nasty behavior in the heavy quark limit
\cite{Brho}. The reason is, when almost the whole momentum is
carried by one of the constituents, the distribution amplitude of
the final meson can be described by the short-distance expansion.
Moreover, the calculation for the form factors is valid only at
the point $q^2=0$, and a pole approximation has to be employed to
study the semileptonic decays. These limit the applicability of
QCD sum rules based on the short-distance expansion of a
three-point correlation function to heavy-to-light transitions and
calls for an expansion around the light-cone, as realized in the
light-cone sum rule approach. In this paper we will try to study
the semileptonic decays of the $B_c$ meson in this approach and
compare the results with the traditional sum rule approach.

The semileptonic decays of the $B_c$ meson involve the transition
$B_c\to \eta_c$, $J/\psi$, $D$, $D^*$, $B$, $B^*$, $B_s$, $B_s^*$.
For the case of $B_c$ decaying into a pseudoscalar meson, to
twist-3 accuracy only the leading twist distribution amplitude
(DA) is involved if we start from a chiral current. If we choose a
suitable chiral current in the vector meson case, the main twist-3
contributions are also eliminated and we can consider the leading
twist contribution only. The result depends on the corresponding
distribution amplitude (DA) of the final meson. We have to
construct realistic models for describing the heavy quarkonium and
other heavy mesons. In particular, the behavior of $\eta_c$ and
$J/\psi$ DA's is an interesting subject by the Belle result for
the cross section $\sigma(e^+e^-\to J/\psi+\eta_c)$. Hence we pay
more attention to discussing the heavy quarkonium DA. We calculate
the cross section $\sigma(e^+e^-\to J/\psi+\eta_c)$ by employing
our charmonium distribution amplitude and the result is consistent
with the experiment data. Based on a phenomenological model for
the leading twist DA, we calculate the form factors for various
$B_c$ decay modes in the corresponding regions. Then we
extrapolate the form factors to the whole kinetic regions, and get
the decay widths and branching ratios for various $B_c$ decay
modes including their $\tau$ modes when they are kinematically
accessible.

This paper is organized as follows. In the following section we
derive the LCSRs for the form factors for various $B_c$ decay
modes. A discussion of the DA models for charmonium and other
heavy mesons is given in section {I}{I}{I}. In section {I}{V} the
cross section $\sigma(e^+e^-\to J/\psi+\eta_c)$ is calculated by
using our charmonium distribution amplitude. Section {V} is
devoted to the numerical result for the semiletonic $B_c$ decays
and comparison with other approaches. The last section is reserved
for summary.

\section{LCSRs for the $B_c$ Semileptonic Form Factors}

~~~According to the definition, the weak transition matrix element
$B_c{\to}P(V)$ can be parametrized in term of the form factors in
the following way:
\begin{eqnarray}
&&{<}P(p_2)|\bar{q}\gamma_{\mu}Q|B_c(p_1){>}=f_+(q^2)(p_1+p_2)_\mu+f_-(q^2)q_\mu,\label{eq:def}\\
&&<V(p_2)|\bar{q}\gamma_\mu(1-\gamma_5)Q|B_c(p_1)>=-ie^*_\mu(m_{B_c}+m_V)A_1(q^2)+i(p_1+p_2)_\mu(e^*q)\frac{A_+(q^2)}{m_{B_c}+m_V}\nonumber\\
&&~~~~~~~~~~~~~~~~~~~~~~~~~~~~~~~~~~~~~+iq_\mu(e^*q)\frac{A_-(q^2)}{m_{B_c}+m_V}+\epsilon_{\mu\alpha\beta\gamma}e^{*\alpha}q^\beta
(p_1+p_2)^\gamma\frac{V(q^2)}{m_{B_c}+m_V},
\end{eqnarray}
where $q=p_1-p_2$ is the momentum transfer,  $e^*_{\mu}$ is the
polarization vector of the vector meson.

For $B_c\to P l\tilde{\nu}$ we follow Ref.~\cite{Bpi} and consider
the correlator $\Pi_\mu(p,q)$ with the chiral current,
\begin{eqnarray}
\Pi_\mu(p,q)&=&i\int d^4xe^{iqx}{<}P(p)|T\{\bar{q}(x)\gamma_\mu(1+\gamma_5)Q_1(x),\bar{Q}_1(0)i(1+\gamma_5)Q_2(0)\}|0{>}\nonumber\\
            &=&\Pi_+(q^2,(p+q)^2)(2p+q)_\mu+\Pi_-(q^2,(p+q)^2)q_\mu.
\end{eqnarray}
A standard procedure, concentrating on $\Pi_+(q^2,(p+q)^2)$,
results in the following LCSR for $f_+(q^2)$:
\begin{equation}
f_+(q^2)=\frac{m_1(m_1+m_2)f_P}{m_{B_c}^2f_{B_c}}e^{m_{B_c}^2/M^2}\int_{\Delta_P}^1{du\frac{\varphi(u)}{u}\exp{[-\frac{m_1^2-\bar
u(q^2-um_P^2)}{uM^2}]}}+\mbox{higher twist terms}\label{eq:ff1}
\end{equation}
with $\bar u=1-u$ and
\begin{equation}
\Delta_P=[\sqrt{(s_0^P-q^2-m_P^2)^2+4m_P^2(m_1^2-q^2)}-(s_0^P-q^2-m_P^2)]/(2m_P^2),\label{eq:delta1}
\end{equation}
where $m_1$ is the mass of the decay quark $Q_1$, $m_2$ the mass
of the spectator quark $Q_2$, and $s_0$ and $M^2$ denote the
corresponding threshold value and the Borel parameter
respectively. In deriving Eq.~(\ref{eq:ff1}) the following
definition of the leading twist distribution amplitude (DA)
$\varphi(u)$ of the pseudoscalar meson has been used:
\begin{equation}
{<}P(p)|T\bar{q}(x)\gamma_{\mu}\gamma_5Q(0)|0{>}=-ip_{\mu}f_P\int_0^1{due^{iupx}\varphi(u)}+\mbox{higher
twist terms},
\end{equation}
with $u$ being the momentum fraction carried by $\bar q$. It has
been pointed out in Ref.~\cite{Bpi} that all the twist-3
contributions have been eliminated so those DA's entering the
higher twist terms in Eq.~(\ref{eq:ff1}) are of at least twist 4.
By repeating the procedure for $\Pi_-(q^2,(p+q)^2)$ we find a
simple relation between $f_+(q^2)$ and $f_-(q^2)$ up to this
accuracy:
\begin{equation}
f_-(q^2)=-f_+(q^2)
\end{equation}

For $B_c \to Vl\tilde{\nu}$ we choose the following correlator as
our starting point:
\begin{eqnarray}
\Pi_\mu(p,q)&=&-i\int d^4xe^{iqx}{<}V(p)|T\{\bar{q}(x)\gamma_\mu(1-\gamma_5)Q_1(x),\bar{Q}_1(0)(1+\gamma_5)Q_2(0)\}|0{>}\nonumber\\
            &=&\Gamma^1e^*_\mu-\Gamma^+(e^*q)(2p+q)q_\mu-\Gamma^-(e^*q)q_\mu+i\Gamma^V\varepsilon_{\mu\alpha\beta\gamma}e^{*\alpha}q^{\beta}p^{\gamma}.
\end{eqnarray}
Also we take the standard definition of the  twist-$2$ and
twist-$3$ distribution amplitudes of the vector meson (see, e.g.,
Ref.~\cite{BC}), and neglect higher twist DA's which is supposed
to be less important in comparison with those written below:
\begin{eqnarray}
&&<V(p)|\bar{q}_\beta(x)Q_\alpha(0)|0>=\frac{1}{4}\int^1_0due^{iupx}\{f_Vm_V[\hat{e}^*g^{(v)}_\perp(u)+\hat{p}\frac{(e^*x)}{(px)}(\phi_\parallel(u)-g^{(v)}_\perp(u))]\nonumber\\
&&~~~~~~~~~~~~~-if^T_V\sigma_{\mu\nu}e^{*\mu}p^\nu\phi_\perp(u)+\frac{m_V}{4}(f_V-f^T_V\frac{m_q+M_Q}{m_V})\epsilon^\mu_{\nu\alpha\beta}\gamma_\mu\gamma_5e^{*\nu}
p^\alpha x^{\beta}g^{(a)}_\perp(u)\}_{\alpha\beta},\label{eq:DA0}
\end{eqnarray}
In Eq.~(\ref{eq:DA0}) $u$ is also the momentum fraction of $\bar
q$, and $m_q(M_Q)$ is the mass of $\bar q(Q)$.

Similarly one can obtain the following sum rules for $
A_1(q^2),A_\pm(q^2)$ and $ V(q^2)$ in Eq.(2):
\begin{eqnarray}
&&A_1(q^2)=\frac{f^T_V(m_1+m_2)}{f_{B_c}m_{B_c}^2(m_{B_c}+m_V)}e^{m_{B_c}^2/M^2}\nonumber\\
&&~~~~~~~~~~~~\int^1_{\Delta_V}\frac{du}{u}\exp{\left[-\frac{m_1^2-(1-u)(q^2-um_V^2)}{uM^2}\right]}\frac{m_1^2-q^2+u^2m_V^2}{u}\phi_\perp(u),\\
&&A_+(q^2)=\frac{f^T_V(m_1+m_2)(m_{B_c}+m_V)}{f_{B_c}m_{B_c}^2}e^{m_{B_c}^2/M^2}\nonumber\\
&&~~~~~~~~~~~~\int^1_{\Delta_V}\frac{du}{u}\exp{\left[-\frac{m_1^2-(1-u)(q^2-um_V^2)}{uM^2}\right]}\phi_\perp(u),\\
&&A_-(q^2)=-A_+(q^2),\\
&&V(q^2)=A_+(q^2)
\end{eqnarray}
with
\begin{equation}
\Delta_V=[\sqrt{(s_0^V-q^2-m_V^2)^2+4m_V^2(m_1^2-q^2)}-(s_0^V-q^2-m_V^2)]/(2m_V^2),\label{eq:delta2}
\end{equation}
and also $m_1$ the mass of the decay quark $Q_1$, $m_2$ the mass
of the spectator quark $Q_2$.

\section{The Distribution Amplitudes of the Charmonium and other heavy mesons}
The leading twist distribution amplitude for the heavy quarkonium,
as defined in the previous section, can be related to the
light-cone wave function $\psi_M^f(x,\mathbf{k}_\perp)$ as:
\begin{equation}
\varphi_M(x)=\frac{2\sqrt{6}}{f_M}\int{\frac{d^2\mathbf{k}_\perp}{16\pi^3}\psi_M^f(x,\mathbf{k}_\perp)}
\end{equation}
where $f_M$ is the decay constant. In the non-relativistic case,
the distribution amplitude $\varphi_M(x)$ goes to the
$\delta$-like function and the peak is at the point $x=1/2$. For
heavy quarkonium, $\eta_c$, the DA should be wider than the
$\delta$-like function since the $c$ quark is not heavy enough. Of
course, it goes to $\delta$-function as the heavy quark mass
$m_c^*\to\infty$.

\begin{figure}[h]
$$\epsfxsize=0.60\textwidth\epsffile{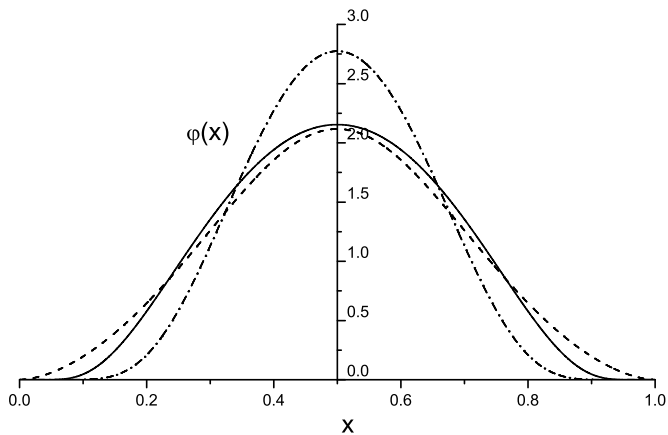}$$
\caption[]{The model for the leading twist distribution amplitude
for $\eta_c$ (in solid line), in comparison with the one in
Ref.~\cite{BC} (dashed line) and Ref.~\cite{eta} (dash-dotted
line). }\label{fig:DA1}
\end{figure}

For the massive quark-antiquark system, Ref.~\cite{harmonic}
provides a good solution
$\psi_{\rm{C.M}}(\vec{q}^2)=A\exp(-b^2\vec{q}^2)$ of the bound
state by solving the Bethe-Salpeter equation with the the harmonic
oscillator potential in the instantaneous approximation. Then one
can apply Brodsky-Huang-Lepage (BHL) prescription \cite{BHL}:
\begin{equation}
\psi_{\rm{C.M}}(\vec{q}^2)\leftrightarrow\psi_{\rm{LC}}\left(\frac{\mathbf{k}_\perp^2+m_Q^{*2}}{x(1-x)}-M^2\right)
\end{equation}
and get the momentum space LC wave function:
\begin{equation}
\psi_M(x,\mathbf{k}_\perp)=A_M\exp\left[-b_M^2\frac{\mathbf{k}_\perp^2+m_Q^{*2}}{x(1-x)}\right]\label{eq:wf1}
\end{equation}
where $m_Q^*$ is the heavy quark mass and $M$ is the mass of the
quarkonium. Furthermore, the spin structure of the light-cone wave
function should be connected with that of the instant-form wave
function by considering the Wigner-Melosh rotation. As a result,
the full form of the light-cone wave function should be
\begin{equation}
\psi^f_M(x,\mathbf{k}_\perp)=\chi_M(x,\mathbf{k}_\perp)\psi_M(x,\mathbf{k}_\perp)\label{eq:wf2}
\end{equation}
with the Melosh factor
\begin{equation}
\chi_M(x,\mathbf{k}_\perp)=\frac{m_Q^*}{\sqrt{\mathbf{k}^2_\perp+m_Q^{*2}}}\label{eq:melosh1}
\end{equation}
After integrating out $\mathbf{k}_\perp$, the leading-twist
distribution amplitude of the heavy quarkonium becomes
\begin{equation}
\varphi^f_M(x)=\frac{\sqrt{6}A_Mm_Q^*}{8\pi^{3/2}f_Mb_M}\sqrt{x(1-x)}[1-Erf(\frac{b_Mm_Q^*}{\sqrt{x(1-x)}})]\label{eq:DA1},
\end{equation}
where $Erf(x)=\frac{2}{\pi}\int^x_0{\exp({-t^2})dt}$. As
$m_Q^*\to\infty$, $\varphi_M(x)$ goes to the $\delta$-like
function certainly. This model of the $\eta_c$ distribution
amplitude has been used to study the large-$Q^2$ behavior of
$\eta_c$-$\gamma$ and $\eta_b$-$\gamma$ transition form factors in
Ref.~\cite{Cao}. The parameters $A_M$ and $b_M^2$ in
Eq.~(\ref{eq:wf1}) can be determined by two constraints on them
completely. One constraint is from the leptonic decay constant
$f_M$
\begin{equation}
\int_0^1dx\int{\frac{d^2\mathbf{k}_\perp}{16\pi^3}\chi_M(x,\mathbf{k}_\perp)\psi_M(x,\mathbf{k}_\perp)}=\frac{f_M}{2\sqrt{6}}\label{eq:constraint1}
\end{equation}
and another one from the probability of finding the $|Q\bar{Q}>$
state in the heavy quarkonium,
\begin{equation}
\int_0^1dx\int{\frac{d^2\mathbf{k}_\perp}{16\pi^3}|\psi_M(x,\mathbf{k}_\perp)|^2}=P_M\label{eq:constraint2}.
\end{equation}
with $P_M\simeq1$ for heavy quarkonium. Inputting the constituent
mass $m_c^*\simeq1.5~\mbox{GeV}$, and the decay constant
$f_{\eta_c}\simeq0.40~\mbox{GeV}$\footnote{The value of
$f_{J/\psi}$ is taken from the leptonic decay of $J/\psi$:
$\Gamma(J/\psi\to
e^+e^-)=(16\pi\alpha^2/27)(|{f_{J/\psi}}|^2/M_{J/\psi})$,
$f_{J/\psi}\simeq0.41\mbox{GeV}$. The one loop corrections
($\sim\alpha_s/\pi$) to the ratio $\Gamma(\eta_c\to
2\gamma)/\Gamma(J/\psi\to e^+e^-)$ indicate that $f_{\eta_c}$ is
slightly smaller than $f_{J/\psi}$ and we take $f_{\eta_c}\simeq
f_{J/\psi}\simeq0.40 \mbox{GeV}$ on average.}, We get the
corresponding parameters for $\eta_c$:
\begin{equation}
A_{\eta_c}=128.1~\mbox{GeV},b_{\eta_c}=0.427~\mbox{GeV}^{-1}
\end{equation}
Then the behavior of the leading twist $\eta_c$ DA can be given
and the comparison with the model from the QCD sum rule analysis
\cite{eta} and the model in Ref.~\cite{BC} is plotted in
Fig.~\ref{fig:DA1}. The moments of these models are given in
Tab.~\ref{tab:moment}. All the wave functions and corresponding
moments are defined at the soft scale $\mu^*\simeq1\mbox{GeV}$.
However, the appropriate scale $\mu$ for the wave functions
entering the LC sum rules will be $\mu\simeq m_b$ for $b$-quark
decays and $\mu\simeq m_c$ for $c$-quark decays with $m_b$ and
$m_c$ the one loop pole masses. Since $\mu$ is not far from
$\mu^*$, this scale dependence can be neglected in our
calculations for simplicity. From Tab.~\ref{tab:moment} it can be
found that the moments of the model (20) is similar to that in
Ref.~\cite{BC}, but much larger than that in Ref.~\cite{eta}.
Obviously the Melosh factor $\chi_M(x,\mathbf{k}_\perp)\to 1$ in
the heavy quark limit $m_Q^*\to \infty$. If we neglect this factor
and integrate $\mathbf{k}_\perp$ from Eq.(\ref{eq:wf1}), we get
the corresponding distribution amplitude which has a much simple
form:
\begin{equation}
\varphi_M(x)=\frac{\sqrt{3}A_M}{8\pi^2f_Mb_M^2}x(1-x)\exp{\left[-\frac{b_M^2m_Q^{*2}}{x(1-x)}\right]}.
\end{equation}
Actually this is just the wave function proposed in Ref.\cite{eta}
based on the QCD sum rule analysis.

\begin{table}
\caption[]{The moments of our model for the $\eta_c$ distribution
amplitude, compared with that in Ref.~\cite{BC} and
Ref.~\cite{eta}.}\label{tab:moment}
\begin{center}
\begin{tabular}{|c|c|c|c|}
\hline
  $<\xi^n>$           & This work & \cite{BC} & \cite{eta} \\\hline
n=2                   & 0.21      & 0.13      & 0.07     \\\hline
n=4                   & 0.053     & 0.040     & 0.012    \\\hline
n=6                   & 0.018 & 0.018     & 0.003      \\\hline
\end{tabular}
\end{center}
\end{table}

For the vector charmonium, $J/\psi$, it is expected that the
behavior of the transverse distribution amplitude is the same as
that of the longitudinal DA since there is no light quark in the
charmonium system, ie.:
\begin{equation}
\phi_\parallel(x)=\phi_\perp(x)=\varphi^f_{\eta_c}(x)\label{eq:DA2}
\end{equation}
which is confirmed by the moment calculation in the QCD sum rules
\cite{Jpsi}.

For the $D$, $B$ and $B_s$ meson, which are composed by one heavy
($\bar Q_1$) and one light quark ($Q_2$), According to the BHL
prescription one takes the following connection:
\begin{equation}
\psi_{\rm{C.M}}(\vec{q}^2)\leftrightarrow
\psi_{\rm{LC}}\left(\frac{\mathbf{k}_\perp^2+m_1^{*2}}{x}+\frac{\mathbf{k}_\perp^2+m_2^{*2}}{1-x}-M_P^2\right).
\end{equation}
with $m_1^*(m_2^*)$ the constituent quark mass of $\bar Q_1(Q_2)$,
$x$ the momentum fraction carried by $\bar Q_1$. Also the Melosh
factor should be modified as
\begin{equation}
\chi_P(x,\mathbf{k}_\perp)=\frac{(1-x)m_1^*+xm_2^*}{\sqrt{\mathbf{k}^2_\perp+((1-x)m_1^*+xm_2^*)^2}}\label{eq:melosh}.
\end{equation}
From which we get the light-cone wave function for pseudoscalar
meson
\begin{equation}
\psi_P^f(x,\mathbf{k}_\perp)=A_P\chi_P(x,\mathbf{k}_\perp)\exp{\left[-b_P^2\left(\frac{\mathbf{k}_\perp^2+m_1^{*2}}{x}+\frac{\mathbf{k}_\perp^2+m_2^{*2}}{1-x}\right)\right]}\label{eq:wf3}
\end{equation}
and the corresponding distribution amplitude\footnote{This model
has been used in Refs.~\cite{BD,BcD} for the $D$ meson
distribution amplitude with the different parameters. There was a
misprint of the factor $\sqrt{2}$ with the decay constant in
Ref.~\cite{BD}.}
\begin{equation}
\varphi_P(x)=\frac{\sqrt{6}A_Py}{8\pi^{3/2}f_Pb_P}\sqrt{x(1-x)}[1-Erf(\frac{b_Py}{\sqrt{x(1-x)}})]\exp{[-b_P^2\frac{(xm_2^{*2}+(1-x)m_1^{*2}-y^2)}{x(1-x)}]}\label{eq:DA3}
\end{equation}
where $y=xm_2^*+(1-x)m_1^*$. Similarly, there are two constraints
Eq.~(\ref{eq:constraint1}) and Eq.~(\ref{eq:constraint2}) to
determine the unknown parameters. We take $P_D\simeq0.8$,
$P_B\simeq P_{B_s}\simeq1.0$ as suggested in Ref.~\cite{wf}.
Inputting the decay constants (We use the least-squares fit values
of the results reported by the CLEO Collaboration \cite{CLEO} and
lattice simulations \cite{MILC, LAT, UKQCD, LAT2, HPQCD, LAT3},
see Tab.~\ref{tab:decayconstants}) and the constituent quark
masses
$m_u^*=0.35~\mbox{GeV},m_s^*=0.5~\mbox{GeV},m_c^*=1.5~\mbox{GeV},m_b^*=4.7~\mbox{GeV}$,
we get the parameters:
\begin{eqnarray}
&&A_D=116~\mbox{GeV}~~~b_D=0.592~\mbox{GeV}^{-1},\nonumber\\
&&A_B=1.07\times10^4~\mbox{GeV}~~~b_B=0.496~\mbox{GeV}^{-1},\nonumber\\
&&A_{B_s}=2.65\times10^4~\mbox{GeV}~~~b_{B_s}=0.473~\mbox{GeV}^{-1}.
\end{eqnarray}

\begin{table}
\caption[]{Leptonic decay constants (MeV) used in the
least-squares fit for our model
parameters.}\label{tab:decayconstants}
\begin{center}
\begin{tabular}{|c|c|c|}
\hline
                  & This work  & other \\\hline
  $f_D$           &    223    & $222.6\pm16.7^{+2.8}_{-3.4}$ CLEO\cite{CLEO} \\\hline
                  &           & $201\pm3\pm17$               MILC LAT\cite{MILC}\\\hline
                  &           & $235\pm8\pm14$               LAT\cite{LAT}\\\hline
                  &           & $210\pm10^{+17}_{-16}$       UKQCD LAT \cite{UKQCD}\\\hline
                  &           & $211\pm14^{+2}_{-12}$        LAT\cite{LAT2}\\\hline
  $f_B$           &    190    & $216\pm9\pm19\pm4\pm6$       HPQCD LAT\cite{HPQCD}\\\hline
                  &           & $177\pm17^{+22}_{-22}$       UKQCD LAT\cite{UKQCD}\\\hline
                  &           & $179\pm18^{+34}_{-9}$        LAT\cite{LAT2}\\\hline
  $f_{B_s}$       &    220    & $259\pm32$                   HPQCD LAT\cite{HPQCD} \\\hline
                  &           & $204\pm16^{+36}_{-0}$        LAT\cite{LAT2}\\\hline
                  &           & $260\pm7\pm26\pm8\pm5$       LAT\cite{LAT3}\\\hline
                  &           & $204\pm12^{+24}_{-23}$       UKQCD LAT\cite{UKQCD}\\\hline
\end{tabular}
\end{center}
\end{table}

\begin{figure}[h]
$$\epsfxsize=0.60\textwidth\epsffile{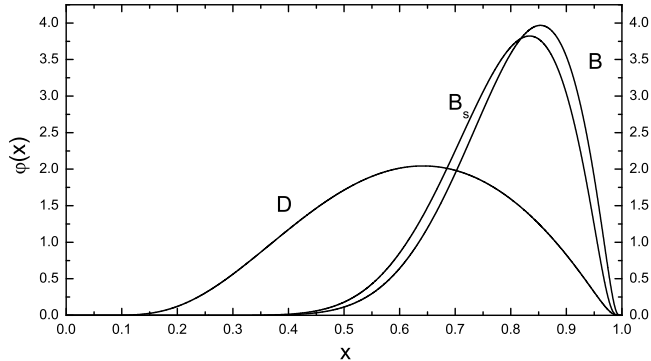}$$
\caption[]{The leading twist distribution amplitudes for
heavy-light pseudoscalar mesons.}\label{fig:DA2}
\end{figure}

The distribution amplitudes of these heavy-light mesons are
plotted in Fig.~\ref{fig:DA2}. The distribution amplitudes of the
corresponding vector mesons are treated in the same way as
$J/\psi$.

\section{The Cross Section $\sigma(e^+e^-\to J/\psi+\eta_c)$}
~~~Following Ref.~\cite{BC}, the cross section $\sigma(e^+e^-\to
J/\psi+\eta_c)$ can be calculated by using the distribution
amplitudes (\ref{eq:DA1}) and (\ref{eq:DA2}). We neglect the
complicated but slow logarithmic evolution of wave function forms,
and account only for the overall renormalization factors of the
local tensor and pseudoscalar currents and for running of the
quark mass, as in Ref.~\cite{BC}. One obtains
\begin{equation}
\sigma(e^+e^-\to
J/\psi+\eta_c)\simeq22.8~\mbox{fb}\label{eq:sigma}
\end{equation}
which is consistent with Belle and BaBar's measurements
~\cite{Belle,BaBar} of this cross section:
\begin{eqnarray}
\sigma(e^+e^-\to J/\psi+\eta_c)&=&25.6\pm2.8\pm3.4~\mbox{fb}\mbox{
  (Belle)},\nonumber\\
\sigma(e^+e^-\to
J/\psi+\eta_c)&=&17.6\pm2.8^{+1.5}_{-2.1}~\mbox{fb}\mbox{
(BaBar)}.
\end{eqnarray}
The value given by Eq.~(\ref{eq:sigma}) is the same order of the
numerical result in Ref.~\cite{BC} and much larger than the
standard non-perturbative QCD (NRQCD) calculation. The reason is
that the DA behavior of the charmonium in our paper and
Ref.~\cite{BC} is much wider than $\delta$-like function due to
the relativistic effect. Also our result confirms the observation
by Ref.~\cite{Ma}. It may be expected that the large disagreement
between the experimental data and the standard NRQCD calculation
can be resolved by combining the light-cone wave function with
relativistic effect and radiative corrections \cite{Zhao}.

\section{Numerical Result for semileptonic $B_c$ decays}
~~~For the decay constant of the $B_c$ meson, we recalculate it in
the two-point sum rules using the following correlator
\begin{equation}
K(q^2)=i\int{d^4xe^{iqx}<0|\bar{c}(x)(1-\gamma_5)b(x),\bar{b}(0)(1+\gamma_5)c(0)|0>},
\end{equation}
for consistency. The calculation is performed to leading order in
QCD, since the QCD radiative corrections to the sum rule for the
form factors are not taken into account. We also neglect the
higher power correction corresponding to the gluon condensates.
The value of the threshold parameter $s_0$ is determined by
requiring the experimental value of $B_c$ be obtained in the
reduced sum rule after taking the derivative of the logarithm of
the SR with respect to $1/M^2$  . The quark mass parameters
entering our formulas are the one-loop pole masses for which we
use $m_b=4.7~\mbox{GeV}$ and $m_c=1.3~\mbox{GeV}$ (cf. Tab. 3 and
Tab. 4 in the review \cite{quarkmass} and references therein). To
get the experimental value $m_{B_c}=6.286~\mbox{GeV}$ \cite{CDF2},
we find $s_0$ should be $s_0\simeq42.0~\mbox{GeV}^2$, which is
smaller than the threshold value input in the ordinary sum rule
\cite{fBc}. This will ensure in some sense that the scalar
resonances will make less contribution in our sum rule. The
corresponding value of $f_{B_c}$ is $f_{B_c}=0.189~\mbox{GeV}$,
which is smaller than that in Ref.~\cite{fBc} since we do not
include the $\alpha_s$ corrections. The same set of parameters
will be used in the LCSRs for the form factor in order to reduce
the quark mass dependance. Take the derivative of the logarithm of
the LCSR for the form factors with respect to $1/M^2$, we get a
sum rule for the mass of the $B_c$ meson . Requiring this sum rule
to be consistent with the experiment value at $q^2=0$, we can
determine $M^2$ for each LCSR. This results in $M^2(B_c\to
\eta_c)=25.8~\mbox{GeV}^2, M^2(B_c\to
D)=11.6~\mbox{GeV}^2,M^2(B_c\to B)=112~\mbox{GeV}^2, M^2(B_c\to
B_s)=111~\mbox{GeV}^2$. It seems that the Borel parameters for
$B_c\to B(B_s)$ are somewhat large. However, the LCSRs are quite
stable in the large region of the Borel parameter
$50~\mbox{GeV}^2<M^2<150~\mbox{GeV}^2$ actually and we just use
the above value for explicit calculation. For the vector meson we
simply use the same $M^2$ as the corresponding pseudoscalar meson
just as we do for the DA's. Also we make the assumption that
$f_V^T=f_V=f_P$.

With all the parameters chosen, we can proceed to calculate all
the form factors involved. The results of the form factors at
$q^2=0$ are given in Tab.~\ref{tab:ff} in comparison with those
from other approaches. Notice that in our calculation we always
have:
\begin{eqnarray}
f_+(q^2)&>&0,f_-(q^2)<0,A_1(q^2)>0\nonumber\\
A_+(q^2)&>&0,A_-(q^2)<0,V(q^2)>0.
\end{eqnarray}
In the 3PSR approach the same relations can be obtained, but only
in the case of nonrelativistic description for both initial and
final meson states, eg., $B_c\to J/\psi(\eta_c)$. In these decay
modes the QM results show the same signature pattern, as can be
seen in Tab.~\ref{tab:ff}.
\begin{table}
\caption[]{The values of the form factors at $q^2=0$ in comparison
with the estimates in the three points sum rule (3PSR) (with the
Coloumb corrections included) \cite{3PSR2} and in the quark model
(QM) \cite{RQM} .}\label{tab:ff}
\begin{center}
\begin{tabular}{c|c|cc|cccc}
\hline\hline
Mode&&$f_+(0)$&$f_-(0)$&$A_1(0)$&$A_+(0)$&$A_-(0)$&$V(0)$\\\hline
                    &   This work      &   0.87  &  -0.87  & 0.75 & 1.69 & -1.69 & 1.69 \\
$B_c\to \bar cc[1S]$&   3PSR~\cite{3PSR2}&   0.66  &  -0.36  & 0.63 & 0.69 & -1.13 & 1.03 \\
                    &   QM~\cite{RQM}    &   0.76  &  -0.38  & 0.68 & 0.66 & -1.13 & 0.96 \\
\hline
                    & This work      & 1.02 &  -1.02  & 1.01 & 9.04 & -9.04 & 9.04 \\
$B_c\to B_s^{(*)}$  & 3PSR~\cite{3PSR2}& 1.3  &  -5.8   & 0.69 &-2.34 & -21.1 & 12.9 \\
                    & QM~\cite{RQM}    &-0.61 &   1.83  &-0.33 & 0.40 &  10.4 & 3.25 \\
\hline
                    & This work        & 0.90 &  -0.90  & 0.90 &  7.9 &  -7.9 &  7.9 \\
$B_c\to B^{(*)}$    & 3PSR~\cite{3PSR2}& 1.27 &  -7.3   & 0.84 & -4.06&  -29.0& 15.7 \\
                    & QM~\cite{RQM}    &-0.58 &   2.14  &-0.27 &  0.60&   10.8& 3.27 \\
\hline
                    & This work      & 0.35 &  -0.35  & 0.32 & 0.57 & -0.57 & 0.57 \\
$B_c\to D^{(*)}$    & 3PSR~\cite{3PSR2}& 0.32 &  -0.34  & 0.43 & 0.51 & -0.83 & 1.66 \\
                    & QM~\cite{RQM}    & 0.69 &  -0.64  & 0.56 & 0.64 & -1.17 & 0.98 \\
\hline
\end{tabular}
\end{center}
\end{table}

Our calculations for the form factors only valid in limited
regions where the operator product expansion (OPE) goes
effectively. For $b$-quark decays, the LCSR is supposed to be
valid in $0<q^2<m_b^2-2m_b\Lambda_{\rm{QCD}}\simeq15~\mbox{GeV}$
and for $c$-quark decays
$0<q^2<m_c^2-2m_c\Lambda_{\rm{QCD}}\simeq0.4~\mbox{GeV}$. It turns
out that the calculated form factors for can be fitted excellently
by the parametrization:
\begin{equation}
F_i(q^2)=\frac{F_i(0)}{1-a_iq^2/m_{B_c}^2+b_i(q^2/m_{B_c}^2)^2}\label{eq:para}.
\end{equation}
Extrapolate the calculated form factors to whole kinetic region
using this parametrization, we can proceed to calculate the
branching ratios of the simileptonic decays of $B_c$. The results
is shown in Tab.~\ref{tab:BR} together with those of other
approaches, where we have used the following CKM-matrix elements:
\begin{eqnarray}
V_{cb}&=&0.0413,~V_{ub}=0.0037,\nonumber\\
V_{cs}&=&0.974,~V_{cd}=0.224.
\end{eqnarray}
For the $b$-quark decay modes in the $B_c$ meson, our results for
the branching ratios are much larger than the corresponding
results in the 3PSR approach. In these decays the kinetic region
is rather large, so the branching ratios depend slightly on the
absolute value of the form factors at $q^2=0$. In the LCSR
approach, the form factors always increase much faster than the
simple pole approximation required in the 3PSR analysis, which
accounts for the discrepancy in these decays. For the $c$-quark
decays in the $B_c$ meson, where the kinetic region is narrow
enough, our results are consistent with the 3PSR approach roughly.

\begin{table}
\caption[]{Branching ratios (in \%) of simileptonic $B_c$ decays
into ground state charmonium states, and into ground charm and
bottom meson states. For the lifetime of the $B_c$ we take
$\tau(B_c)=0.45\rm{ps}$.}\label{tab:BR}
\begin{center}
\begin{tabular}{c|cccc}
\hline\hline
  Mode           &This work & 3PSR~\cite{3PSR2} & QM~\cite{RQM}& ~\cite{BSE}\\\hline
$\eta_ce\nu$     & 1.64 & 0.75 & 0.98 & 0.97  \\
$\eta_c\tau\nu$  & 0.49 & 0.23 & 0.27 & ---   \\
$J/\psi e\nu$     & 2.37 & 1.9  & 2.30 & 2.30  \\
$J/\psi\tau\nu$  & 0.65 & 0.48 & 0.59 & ---   \\
$De\nu$          & 0.020&0.004 & 0.018& 0.006 \\
$D\tau\nu$       & 0.015&0.002 &0.0094& ---   \\
$D^*e\nu$        & 0.035& 0.018& 0.034& 0.018 \\
$D^*\tau\nu$     & 0.020& 0.008& 0.019& ---   \\
$Be\nu$          & 0.21 & 0.34 & 0.15 & 0.16  \\
$B^*e\nu$        & 0.32 & 0.58 & 0.16 & 0.23  \\
$B_se\nu$        & 3.03 & 4.03 & 2.00 & 1.82  \\
$B^*_se\nu$      & 4.63 & 5.06 & 2.6  & 3.01  \\
\hline
\end{tabular}
\end{center}
\end{table}

\section{Summary}
The semileptonic decays of the $B_c$ meson are studied in the
Light-Cone sum rule approach. By using suitable chiral currents,
we derive simple sum rules for various form factors, which depend
mainly on the leading twist distribution amplitude of the final
meson. A model with the harmonic oscillator potential for the
light-cone wave function is employed. Special attention is payed
to the leading DA of the charmonium. It has been found that our
model is consistent with the QCD sum rule analysis. Also, the
moments are found to be similar to the model proposed in
Ref.~\cite{BC}. Based on this model, we calculate the form factors
for various $B_c$ decay modes in the corresponding regions.
Extrapolating the form factors to the whole kinetic regions, we
get the decay widths and branching ratios for all the $B_c$
semileptonic decay modes. For the $b$-quark decay modes in the
$B_c$ meson, where the kinetic regions are quite large, our
results for the branching ratios are much larger than the 3PSR
results. For the $c$-quark decays in the $B_c$ meson, they are
consistent with each other in general.

It is a crucial point to construct a realistic model for the light
cone wave function of the charmonium which is not a
non-relativistic subject. Based on the solution of the
relativistic Bethe-Salpeter equation in the heavy quark system, we
provide a model in this paper by using the BHL prescription and
the behavior of the charmonium DA is much wider than $\delta$-like
function which was employed essentially by the approximation of
NRQCD. Thus the cross section $\sigma(e^+e^-\to J/\psi+\eta_c)$
can be enhanced considerably and is about $22.8~\mbox{fb}$.

\newpage

\begin{center}
{\bf ACKNOWLEDGEMENTS}
\end{center}

This work was supported in part by the Natural Science Foundation
of China (NSFC). We would like to thank Dr X. G. Wu for helpful discussions. \\

\end{document}